# From zero to figure hero

A checklist for designing scientific data visualizations

**Helena Klara Jambor**[1,2]

1 – Institute for Data Analysis, Artificial Intelligence, Visualization and Simulation, University of Applied Sciences of the Grisons/Graubünden, Chur, Switzerland

2 – Medical Faculty Universitätsklinikum Carl Gustav Carus an der Technischen Universität Dresden, Germany

Email: helena.jambor [at] fhgr.ch

## Abstract

Biological research spans scales and methodologies, generating complex data visualizations such as images, text, numbers, networks, and maps. With increasingly large and multimodal datasets, effective visualization is essential for efficiently conveying scientific insights. Despite this crucial role, biologist often lack training in data visualization and information design. This work addresses this gap by providing a framework for creating clear, accurate, and impactful visualizations of biological data. It is centered around a checklist that guides biologists through the process of developing publishable figures. The guide and checklist cover key aspects such as selecting appropriate display types, using color palettes effectively, and optimizing figure layouts to communicate complex data. Additionally, the work is supported by evidence from visualization research, ensuring that the checklist recommendations are grounded in established principles. By following this guide, biologists can enhance their visual data presentations, ultimately increasing the impact of their scientific findings on diverse audiences.

## Introduction

Biological research traverses scales, from molecules to systems, and employs chemical, physical, clinical and sociological methodologies. The resulting data are communicated in figures of images, text, numbers, networks, maps and more. Today, biologists communicate their data and insights with the over 1 million papers published each year, each containing numerous data figures (Lee P, 2018). Data figures decrease the time it takes audience to understand the scientific evidences, recognize patterns and trends, and reach conclusions (Ware, 2004).

Image figures were already widespread in antiquity, when natural scientists documented specimen in atlases used for pharmacological and medical applications. Charts and figures for numerical data were pioneered by William Playfair, who developed and popularized pie and line charts in the 18th century for economic data. Subsequently, the process of transforming numbers into visualizations, the `graphical method`, was then introduced in life sciences. John Snow added data to maps to interrogate epidemiological patterns of cholera hot spots in London. Florence Nightingale used graphs to illustrate the undeniable health benefits of introducing hygiene measures in military hospitals. She famously conceived the "rose diagram", which presents data in circular design to reveal periodicity in annual patterns. In 1885 the graphic method, including instructions for presenting images, was introduced to physiology and biology with a first textbook in French (Marey, 1885). Today, developing visualizations for exploring and presenting multidimensional, multiscale, and multimodal biological data has blossomed into a multi-disciplinary research field of computational, bioinformatics, and biological scientists.

Yet, despite regularly working with large and complex data, biologists are rarely trained in the significant task of condensing this complex information into static, two-dimensional figures that are effective in communicating insights to

# From zero to figure hero

Effective figures communicate scientific insights. This checklist helps when designing, improving, and reviewing figures for publications. Before starting, ensure that the key message and target audiences are defined. The checklist can be used in any order suitable to your process and should be consulted iteratively as necessary to create impactful data visualizations.

## Feedback

### The 1-second test
- Ask 2-5 people: What do you see first? Evaluates if data is visible at first glance

### Reverse feedback
- Ask: Explain to me what you see. Get feedback on chart type, text, layout, colors

### Focus the attention
- Axes, boxes, tick marks. Remove or mute?
- Legend necessary? Direct data label or title possible?
- Gridlines. Remove or mute? Necessary for log-scales/ precise values
- Use color sparingly. Remove unnecessary colors, use grey instead of black
- Align chart elements and multi-panel figures ticks, text, titles, legends, axes, labels
- Aim for symmetry, evenly filled space, separate elements in multi-panel figures gaps/white space

## Basics 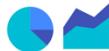

### Choose chart
- Suitable for data and message?
- Suitable for audiences?
Consult resources for chart types:
Datavizcatalogue.com
Python-graph-gallery.com
Datavizproject.com

### Simplify charts
Too much data:
- Split data across 2 charts
- Consider small multiples for many categories/ observations
- Animate information stepwise
Uncommon chart type:
- Include help on how-to-read e.g., in title, subtitle, legend,…
- Use intuitive guides e.g., direct data labels, color similarity, regions-of-interest

### Text in charts
- Label axes
- Label tick marks, choose easy intervals
   good: 0, 5, 10; poor: 0, 7, 14
- Explain colors, marks, shades e.g. legends
- Use title, subtitle to orient readers
- Avoid abbreviations if necessary: test
- Typography: choose legible font and style regular > bold, italic

## Design 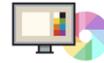

### Layout
- Aim for horizontal text, avoid text rotation
- Label data directly. Otherwise place legend close to data
- Align text elements title best: top left/center
- Use white space to visually separate panel elements

### Encode data with color, color schemes
- Are all colors needed?
- Color palettes for data. Sequential data: 1 color, vary saturation; multi-colors necessary: palette with homogeneous lightness, e.g., viridis (not jet/rainbow). Diverging data: -2 colors diverging from central/neutral point in white/grey; Categorical data: several colors possible.
- Use consistent color schemes
- Ensure colors are accessible. Color-blind safe, high contrast foreground/background. Double encode color information. Best, information is also visible in greyscale.
- Explain all colors e.g., legend

### On beauty
- Align and organize with a grid
- Use white space to separate elements, but do not leave gaps in multi-panel figures
- Strive for symmetry
- Consider using icons. Sources:
General: Nounproject, SVGrepo,
Bio: Bioicons, Phylopic, SciDraw
Medicine: Healthicons, SmartServier

## Specials 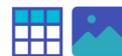

### Tables
- Left align text, right align numbers, align each column header with its content.
- Use font with same width for all numbers
- Eliminate unnecessary cell borders to gain space in cells
- For overview, consider color-coding numbers heatmap

### Image data
- Select a suitable image frame
- Add scale bar with dimensions if possible
- Explain colors scale for quantitative data
- Ensure colors are color blind safe
- Explain annotations symbols, arrows, regions

*Figure 1. Data Figure Checklist*

broad audiences. Edward Tufte's influential books on information design have been seminal in shaping the field of data visualization (Tufte, 2011, 1990). While Tufte's work has set the standard for clarity, precision, and efficiency in visual communication, some of them may not be fully applicable to the specific challenges of visualizing scientific data, where the complexity and detail required often demand a more nuanced approach. A few short articles advised biologists on specific formats, such as designing conference posters (Brown, 1996), which is also covered in a more recent book (Faulkes, 2021). Recently, across sciences but in particular in the life sciences, the focus has shifted to large scale experiments producing even larger datasets. This sparked renewed interest in the possibilities of interrogating and interpreting data also with visualization techniques. Addressing this new need, Nature Methods over several years ran a Points of View series by Bang Wong; the series touched on aspects such choosing display types for data, but also figure layout, color schemes, overview figures, and arrows. Two books guide biologists, one introduces the entire data workflow for an experimental cell biologists, from measurement to figure (Royle, 2019), and another introduces data visualization with R with numerous examples (Wilke, 2019).

## Figure Improvement process

Here, I have distilled key lessons from data visualization into a practical checklist that helps biologists to quickly create impactful data figures (Figure 1). Whether intended for a manuscript, grant proposal, poster presentation, or slide deck, this checklist may be a valuable tool for those involved in figure design—whether this is creating the figures, advising colleagues, reviewing, or teaching figure design. It draws upon insights from numerous courses teaching life science data visualization and extensive consulting experience with scientists.

Before beginning the design process, it's crucial to clearly define your key message and identify your target audience. If the central message is still unclear, the resulting figure is likely to be ambiguous as well. Understanding the target audience—their expectations, experience level, and needs—enables you to tailor the complexity and detail of the data visualization to better communicate with them.

The checklist may be used as a workflow to iteratively refine a figure or consulted in any order according to your personal preference or the specific needs of a project. Whether you use it from the outset of figure creation or as a reference during the process, the ultimate goal is to produce clear, self-explanatory data figures. For each checklist item some background on visual perception and design is provided. Additionally, I have compiled further useful resources and tools (Table 1) and further reading materials (Table 2).

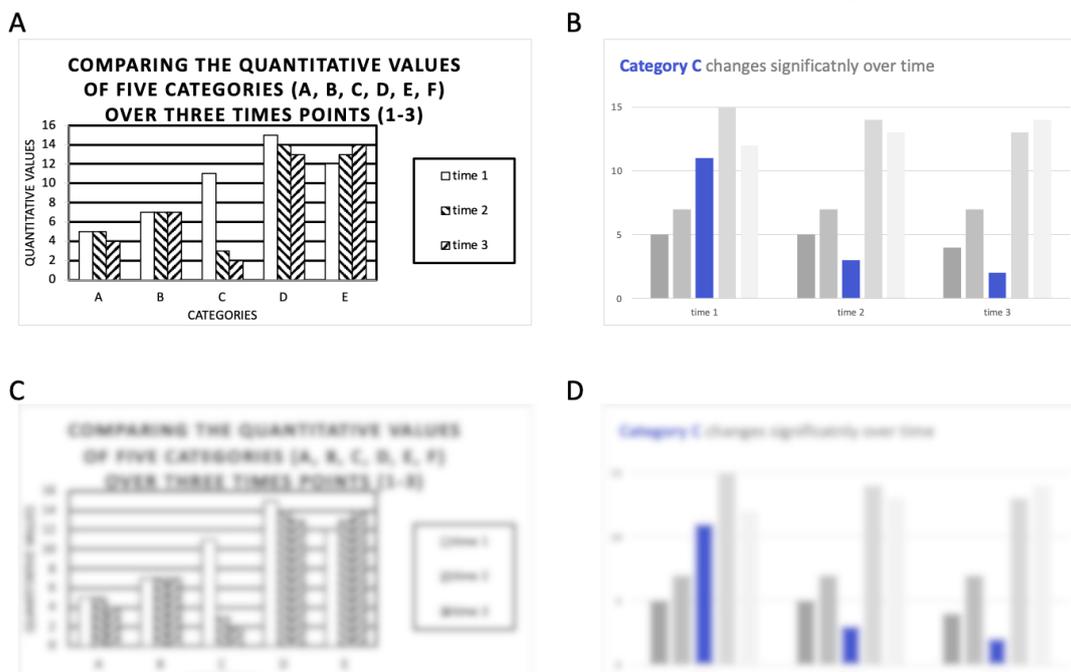

*Figure 2 The same data shown in a basic chart (A) and improved version (B) where color, lines, and text are designed to orient audiences. In the 1-second test the legend, title, gridlines and patterned bars compete for visual attention in the basic chart (C), while the colored words and bars rapidly focus the attention in the improved version (D).*

## The 1-second test

To get very quick impression on what works in a figure, starting with the 1-second test is a good approach. For the 1-second test, ask two or three people to simply tell you what they see first, where do their glances go first. Note, do not ask about they understand the data, just what in the figure draws their attention, what is most noticeable to them at first glance. In design, this is called the "visual weight" or "salience". The attention should be drawn to the main chart or data point, but may accidentally be directed to the brightest color, a weird shape, or a unproportionally large legend (Compare Figure 2). The 1-second test may be already done with draft versions (Figure 2A), and repeated as the figure reaches a final stage (Figure 2B). Early and rapid feedback from the 1-second test allows to assess if a figure is already functional, or needs improving. The test can be done with pretty much anyone who is not too familiar with the data, and thus may offer a fresh point of view, and in fact this does not require a scientific background. I advise to not ask direct collaborators who may immediately be drawn to an interesting data point rather than provide first glance feedback.

*Ask: Where do you look first?*

<u>Background.</u> Human visual perception is limited. This was described by Miller in 1956 as the "Magic number 7" rule, observing that humans can maximally perceive 7 (plus or minus 2) visual elements at a time; today, the limited capacity of our visual perception is discussed as "chunking" in the psychology in learning (Miller, 1956). Human vision is also biased towards certain features, so called preattentive attributes, that catch our visual attention first (Wong, 2010a, 2011a). Attributes are for example color, size, orientation, or shape – the largest object, the brightest color, the outlier datapoint, or a weirdly shaped cell among many spherical ones. To efficiently and effectively transmit information in figures, it is therefore important that those objects with most salience in a figure are also the most relevant (Canham and Hegarty, 2010; Hegarty et al., 2010).

## Reverse feedback

In reverse feedback the goal is to observe someone else trying to figure out a visualization without guidance. The person providing feedback is asked to explain the figure back to its author, while verbalizing their thoughts (Figure 3). The key aspect is that the author refrains from providing any explanations, instead they should merely listen and learn.

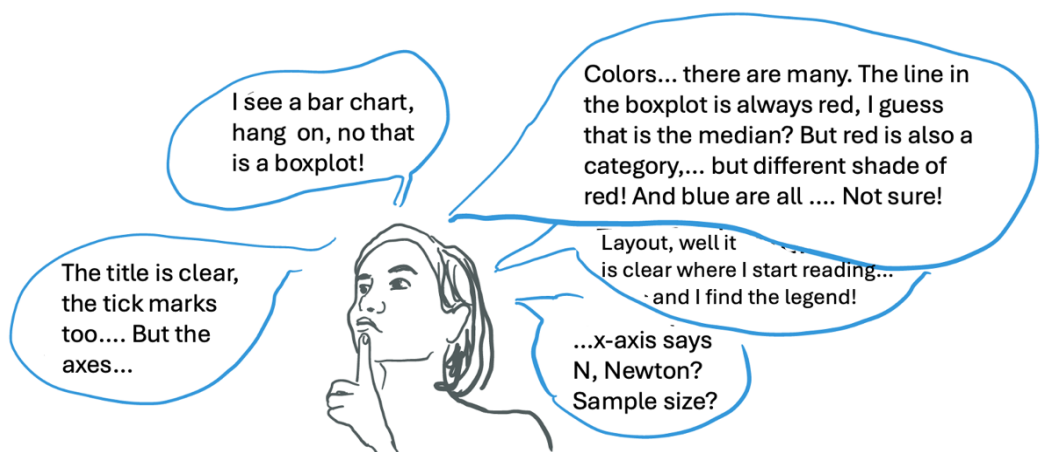

*Figure 3 Example of reverse feedback: the person providing feedback simply verbalizes their attempt to understand the figure without any introduction to it from the author – mimicking a reader of a manuscript who also does not get upfront explanations by the author.*

This setting provides the author with a unique view how audiences interact and interpret the figure. For high quality feedback, it is advisable to seek feedback from someone not familiar with the work itself, but comfortable with scientific figures.

*Ask: Explain to me what you see.*

After a few minutes you can introduce more specific questions (Figure 3):

- Do you understand the data display? (chart, image, table, map, schematic)
- Is the text legible and helpful? (abbreviations, typography)
- Where do you start to read? (layout)
- How do you interpret the colors?

*Background.* In computer sciences evaluations are common in interface design (REF: B. Shneiderman. Designing the User Interface: Strategies for Effective Human-Computer Interaction. Addison-Wesley, Reading, MA, 1987). Several approaches have been established to probe the user interaction with a display. The methods include laboratory observation or eye tracking, A/B testing to compare different versions, and questionnaires, but the first evaluation most often is expert feedback. Expert feedback describes the consultation with professionals who have knowledge and experience with the problem at hand and with best practices and usability issues. Reverse feedback on figures thus is a form of expert feedback (Tory and Moller, 2005). A more formalized version of reverse feedback is heuristic evaluation, where experts review the interface using a defined set of usability principles, so called heuristics, to identify potential issues (Zuk et al., 2006). An example would be when the figure checklist (Figure 1) is used to standardize the evaluation of figures.

## Focus the attention

The feedback from the 1-second test helps to identify features that can potentially be muted or removed to focus the attention. Anything that was noticeable at first glance but is not the central point of the chart should be deemphasized (Figure 4). Like in the example (Figure 2A) this often include long titles, excess annotations, superfluous legends, axis marks in small increments, or decorative and cluttery elements and patterns. Alternatively to removing chart elements, they may also be muted to reduce the visual attention. Muting can be achieved with e.g. smaller point size for text, grey instead of black text, or a thinner version of a fonts (light, non-bold, non-capitalized, see Figure 2B). In general, everything that constitutes as preattentive attribute should be carefully chosen: line widths (e.g., bold, regular and italic font), symbols (e.g., circle filled, circle open, rectangles), and colors, including patterns (Figure 4).

Remove of mute (de-emphasize):

- Boxes, strong axes, abundant tick marks. Many programs default to including bounding boxes to chart elements and axes and tick marks in black and thick stroke. These however should only visually orient audiences, not draw attention – and therefore often can be removed altogether, or at least muted (smaller point size, grey font, no bolding, thinner stroke, fewer marks etc).
- Gridlines. Many charts by default come with gridlines – often they can be removed, reduced to major units, and sufficient in thin, grey stroke. They are usually necessary to signal the use of log-scales.
- Axis labels should be concise and not contain redundancies. Often (parts of) axis labels can be integrated in figure titles, especially the y-axis label (measured quantity).
- Carefully assess which parameter is shown on which axis. The invariable parameter, e.g. observed time, usually is plotted to the x-, the variable/measured parameter to the y-axis; however there are subject specific exceptions.
- Legends are automatically included in many plot types, but are at times not necessary, for instance when the data is also directly labelled.
- Plot backgrounds (e.g. in grey) are often included by default, but invariably reduce the visibility of the data points in the foreground. Choose a background color with high contrast to the data, or remove entirely.

- Colors. Do a count, how many colors are used in the chart and critically assess if they are necessary and visible to all audiences (#2.8)? The rule of thumb, Magic Number 7+/-2 (Miller, 1956) suggests we should limit visual information to seven elements, and this also applies to color, where color blind accessibility further limits the colors to 4-5.

Removing and muting elements in a figure often suffices to focus the attention on the main point of a figure (Figure 2A,B). Another effective method to focus the visual attention is to improve the layout, or, as Tufte would say, "declutter" the design by improving the following elements:

- Ensure consistent alignment of ticks, text, titles, legend elements, and labels within each chart. In multi-paneled figures, align the charts, titles, and panel labels across all panels.
- In multi-paneled figures, aim for symmetry by evenly filling the figure area, ensuring the overall layout ideally forms a rectangle. Eliminate any empty spaces or holes within the grid (figure area) and maintain sufficient white space to separate the panel elements.

The effect of these measures may be tested with another round of the 1-second test, preferentially with new targets, or by soliciting a more detailed feedback, see #3.

*Background.* How we process visual information and forms is already described in the "principles of form" by Bauhaus adjacent designers (Wertheimer, 1912). Later, these principles were broken down into the simplest elements our eyes see, summarizes as the "pre-attentive attributes" (Bertin, 1967). In the 1950ies psychologists also formalized concepts on learning and information processing, to which a key insight are the limits of our perception and later that information is processed in groups (chunks) rather than individual elements (Johnson, 1970; Miller, 1956). Hubel and Wiesel studied the neurobiology of the visual system, which confirmed that basic cues such as color, orientation, size or motion are processed by neurons for in the visual cortex, while complex juxtapositions (e.g. charts) are integrated in a later stage of the brain (Barlow, 1982). Thus by focusing the attention of our audiences on key data rather than unimportant details and the principle of salience, information designers like Tufte adjust visualizations to our neurobiological capabilities (Tufte, 2011).

## Choose a chart

Is paramount to choose a suitable display type for your data. For microscopy data, these are usually a digital image; for explaining a finding, it might be a graphical abstract (Jambor and Bornhäuser, 2024); for numerical data, there are however hundreds of chart types to choose from (Table 1). Selecting a chart type that is both suitable for your data as well as your audiences is an arduous task. Audiences learn the basic chart types and their applications, bar, line or pie chart and maps, in school. When showing trends over time, time is mapped to the x-axis; pie-charts show percentages and should add up to 100; in most maps, North is on top, South at the bottom etc.

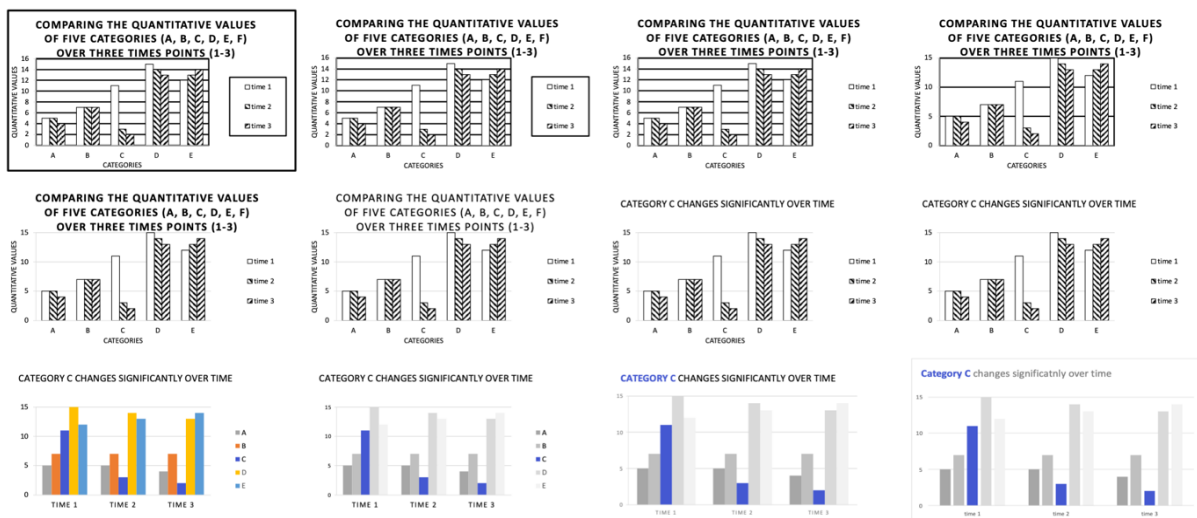

*Figure 4 Possible workflow for removing or muting part of a chart, starting (top left) with unnecessary boxes, to changing the data organization, and using color only to highlight most important data (bottom right).*

Specialized and domain-specific charts are usually learned in advanced degrees, and thus can be read by audiences with similar training, but often not by experts from other domains.

There is also mounting research in computer science for how information is best designed to be decodable by our visual system, and this may be helpful to consider before choosing a chart type. When selecting a chart, be mindful if the intended audiences have the training to understand chosen visual display and that you followed the common practices of using that chart type.

Select a chart type:

- DataViz Catalogue (by Severino Ribecca, http://www.datavizcatalogue.com/)
- Python Graph gallery (by Yan Holz https://python-graph-gallery.com/)
- Chart overview (by the Financial Times https://github.com/Financial-Times/chart-doctor/blob/main/visual-vocabulary/README.md)

*Background*. Information design and how charts work is an actively researched in the computational sciences. Seminal work demonstrated that humans are good in making quantitative assessments values are positioned along a common scale such as bar (comparison of bar height) or scatter plots (comparison of x- and y position), and this task is harder when a common axis is missing like in line and pie charts (Cleveland and McGill, 1984). In many chart types, other tasks are performed: line charts require comparison of direction and angle, maps and heat maps of color or texture, and bubble charts of area (Cleveland and McGill, 1984). For judging area the shape is critical and circles are generally harder to compare than rectangles. However, there is also a hierarchy among rectangles, with squares and rectangles with extreme ratios being hard to interpret (Kong et al., 2010). The effectiveness of tasks was directly compared by presenting the same data as pie, line or bar chart (Hollands and Spence, 1992). They showed that for comparing trends, separating the data into several pie charts for each observation was less effective than having the information grouped in one charts with many bars, where an imaginary line could be drawn to compare bar heights, was similarly effective to a line chart. More recent work is dedicated to the effectiveness of charts for large datasets and distributions. For example, bar charts showing the median and the spread of a distribution are hindering an audience to clearly comprehend how datapoints contribute to the displayed summary statistics, and thus lack transparency (Weissgerber et al., 2015). A comprehensive overview of the science of data visualization was recently published and is a strongly recommended read (Franconeri et al., 2021).

### Simplify charts

You may have chosen a suitable chart for your data, but it seems too complex for effective communication.

Scenario 1, the chart type is in principle easy to read (e.g., bar, line, pie chart), but the message complex. This might be a case for → focusing the attention (Figure 3), improving the → text, → colors, → layout, or simply having too much information. In case of the latter, an quick fix is to split the information into two or even many plots (Figure 5A, Figure 3). This is useful if you want to point out an overarching trend over many items, and still want audiences to see each item separately; compare trends of two variables with a double-axis; show a scatter plot with three or more variables.

Many times, more charts are better than one complex chart due to the limitation of our visual system (Magic number 7). A complex chart with 16 categories can be also shown in a 4x4 grid, each chart showing just one category while allowing comparisons (Figure 3). Such a series of charts with the same design (same scale, same axes) are viewed as grouped, effectively reducing the mental load. Additionally, in a series each chart may be shown at much smaller size and thus can be displayed in the same space as one complex chart.

Check if:

- splitting the message to two charts is possible
- small multiples is an option for > 2 messages
- in talks: animating the chart and successively adding more information makes it easier for audiences.

Scenario 2, the chart type is rather uncommon, but necessary here. In this case it would be helpful to include explanations on how to read the chart such as helpful titles, subtitle, legends, or an annotated chart as legend. As a rule of thumb, there is no chart type that is off limits or impossible to use, even highly unusual charts may communicate data well, provided that they are explained. Check if you can include a how-to-read or subtle helps for your audiences, or include animations (Figure 5B).

Think about:

- a title that explains the chart right away ("Manhattan plot showing the frequency of reads in height for the entire chromosome 21.")
- a title that includes key legend items ("Manhattan plot showing the frequency of reads in **wild-type** and **mutant** samples.")
- direct data labels and annotations

*Background*. Not much research addresses simplifying charts for communication, however visual designers have put forward a number of good practices for simplifying visual information. Splitting information into a series of charts was promoted by Edward Tufte in his landmark books, where he introduced it as "small multiples" (Tufte, 2011). Small multiples can be easily realized in programming software used for data visualization such as Python and R, where the feature is available as gridding and faceting respectively.

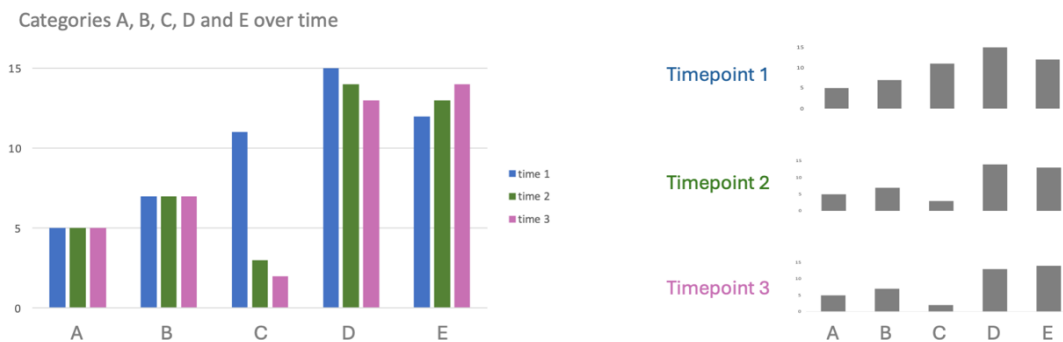

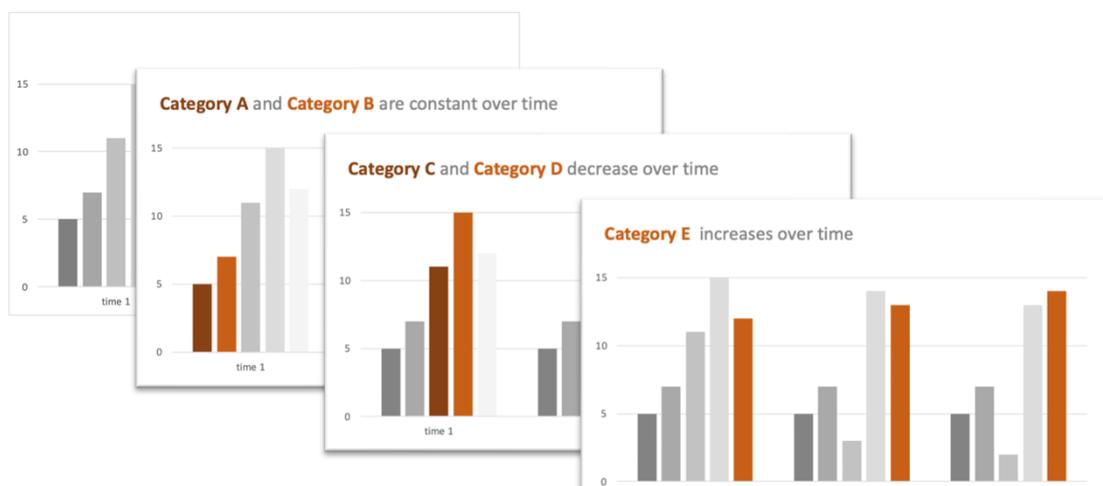

Figure 5 *Strategies for simplifying information in charts. A. Breaking up complex information into several charts with shared axis layout. Note, this usually does not require more space. B. Animation of information in complex charts.*

Among the research articles, one study explored how dense data can be in graphics while still being effective. In the study, positive and negative values in a line chart were collapsed over the x-axis and displayed side by side, distinguishable by their color. They showed that at ¼ of the size, the data was still legible (Heer et al., 2009). An even denser version than the examples studied, are the climate stripe charts, popularized by Ed Hawkings (Hawkins, n.d.).

## Text in charts

When contemplating figure design, text does not immediately is at the forefront of our thoughts. However, every figure includes a lot of text, without which it would be abstract art rather than data information (Figure 6A). Text explains the units, colors, data points, and often links to further information. In order to be helpful, text in figures has to be legible by choosing a sensible font type, style and size, but also comprehensible, meaning it must match the audiences vocabulary. A common pitfall are abbreviations, that biologists are prone to and may use every day in their laboratory, but outside of it quickly become ambiguous. A popular exercise in my workshops involves writing an abbreviation on paper and having 5-10 participants interpret it – only if the majority decodes it correctly, it can be considered suitable (Figure 6B).

Check:

- Do all axes have labels?
- Are tick marks sensibly labelled?
- Is a legend present to explain all colors, marks, and shades?
- Did you include a title and/or figure legend? The better written, the more easy it will be for audiences to get your point!
- Avoid (and if you must use them test) abbreviations. Really only use those your reader are familiar with. Often only specialist audiences know the numerous acronyms of your subject. When aiming for high impact with broad audiences, only use very common abbreviations (e.g., DNA, RNA. Figure 6).
- Generally, avoid mixing font and font styles for emphasis. It is common in advertisements, but in scientific work readers are often puzzled and look for cues to its meaning. Bolding reduces the legibility by lowering the white space around letters, italics by changing the direction of ascenders and descenders.

*Background.* Complex visual insights are most memorably when they effectively integrate text and include keywords and phrases known to the audiences (Hegarty and Just, 1993; Martin, 2020). Indeed, a study quantified a link between the citations of an article and understandable journal titles that are concise, devoid of jargon, and limited to common abbreviations (Letchford et al., 2015). At times it can increase clarity as well as memorability when text is supplemented with visuals such as icons and pictograms (Haroz et al., 2015). Typography, i.e. which font is chosen, is an important aspect for graphic designers, usually scientists however are restricted by publishers, grant agencies, and universities corporate design formatting requests. If the opportunity arises to choose a font, ensure that a legible font is used (Wong, 2011b). There is no consensus on whether a serif (e.g., Times New Roman) or sans serif (e.g., Arial) is more readable (Arditi and Cho, 2005). A legible font has a constant height of all letter bodies (x-height), visible ascenders (e.g., in b) and descenders (e.g., in p), and enough white space in closed letters (e.g., a, o).

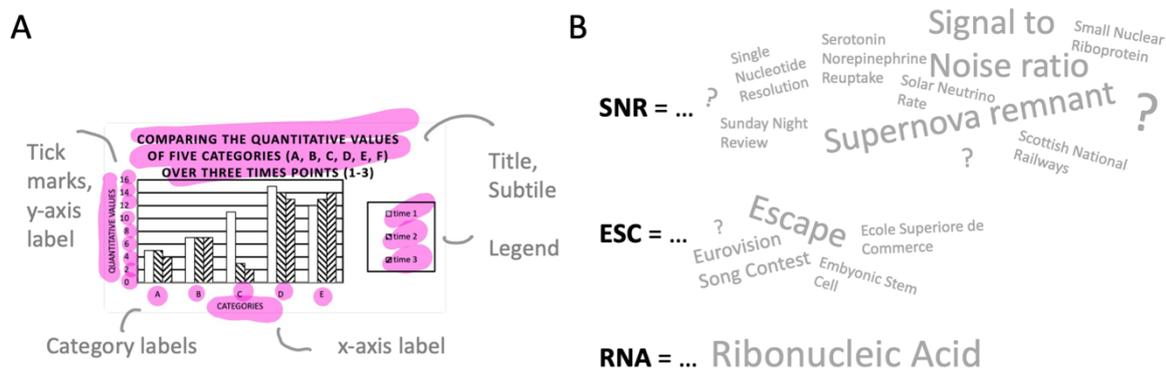

*Figure 6 A. All charts come with text and labeling. B. Polling scientists about the meaning of an abbreviation, with the most common responses displayed in larger font text. Examples and their intended use: SNR (microscopy and physics: signal to noise ratio). ESC (embryonic stem cells). RNA (ribonucleic acid).*

### Layout

When mentioning layout, most scientists think of figure with panels from A to P, or of scientific posters. These require a clear reading direction, organization on a grid and enough white space to unambiguously orient the reader. However, also within a chart authors should consider the layout to increase legibility. This involved deciding where to place text (top left for title and bottom for e.g. data sources is easiest to design), whether it is organized in horizontal (good for reading) or vertical (may reduce required space, but also readability) orientation, and whether a legend is required or data can be directly labelled. For the chart itself, oftentimes authors are able to choose which data is added to which axis, and how wide an axis will appear.

Be attentive:

- Avoid text rotation, every degree of rotation is slowing down our reading speed and, quite honestly, annoying. If possible, keep all text horizontal.
- Direct data labels, a secret power to effective charts that is underused.
- Meticulous alignment: in multi-panel figures, organize elements on a grid. In single-chart figures carefully position title, subtitle, axis labels
- Use enough white space (negative space) to clearly separate elements, but not leave gaps in multi-panel figures.

*Background*. Further resources on how to organizing figure panels (Wong, 2011c, 2011d), graphical abstracts (Jambor and Bornhäuser, 2024) and posters (Brown, 1996; Faulkes, 2021; Jambor, 2023) is available, as well as tools that help authors quickly arranging items in a figure panel (e.g., in ImageJ: Aigouy and Mirouse, 2013; Mutterer and Zinck, 2013).

### Encode data with color

Quantitative data is not only displayed by length (bar chart), area (pie chart), or angle (line chart), but can also be encoded with color. Color is used to visualize groups in the data, e.g. different categories in a scatter plot, or show measurement values, e.g. in a heatmap. Depending on the data, different color schemes are suitable (Figure 7).

For categorical data (also referred to as "qualitative") we vary the color, less often the pattern (e.g. in maps) or the symbol (e.g., in scatterplots). We collect categorical data when we measure different species in zoology, different treatment groups in clinical trials, or different proteins in biochemistry.

For sequential data (also referred to as "continuous") we use color palettes with a single hue that varies in saturation (light blue to dark blue). We collect sequential data when we measure gene expression, the size of specimen, or reaction rates. A special case are sequential data that requires a multi-color scale for better differentiation. This is often used when data has a topological distribution, e.g. a physiological gradient in a cell, geographical density of birds, or cell marker measurements in a FACS scatterplot.

A third category is sequential data that is however diverging from a central point – this may be a natural point of divergence (zero, sea level, 0°C, pH 7) or a calculated one (above or below mean, threshold, or normalized value). For diverging sequential data, we use a two-color palettes with white or grey representing the central values.

Make sure:

- Are all colors needed? If not, leave out!
- Categorical data: vary color, this reflects that there is no inherent order in the categories.
- Sequential data: one color, vary saturation or lightness. For special cases: multi-color palettes which sequentially vary in lightness (e.g., viridis).
- Diverging data: two colors, sequentially joined through white or grey for the center data.
- Include explanations for colors used, e.g. legends.

*Background*. "Color" is describing a mixture of three parameters, the hue (red, green), its saturation (bright red, pale red), and its lightness (dark-light) (Wong, 2010b). Color belongs to the core preattentive attributes as it has strong and immediate effects on visual perception. Thus, when color is used to encode quantities care must be taken to select a color palette suitable and truthful to the data it encodes (Gehlenborg and Wong, 2012; Hattab et al., 2020; Szafir, 2018). This is especially important when sequential colors are encoded with multi-color palettes. Oftentimes, jet or rainbow color palettes were used for this case, however these vary both the color as well as the lightness – the result are inhomogeneous perceived color changes (border effects), that do not faithfully represent the sequential chances in the data. Addressing this issue, new multi-hued palettes were developed, such as Viridis, which do not vary in lightness (Smith and van der Walt, n.d.). A useful resource to quickly orient how to use color to encode data was developed by Cynthia Brewer, originally for cartographic data but widely used across all scientific domains today (Brewer, n.d.; Harrower and Brewer, 2003).

## Color schemes

Regardless of the data type, all colors should be chosen with accessibility in mind and should be used consistently across e.g. all figures of a paper. Accessibility means that they are visible to color-blind audiences and also provide good contrast with the background to allow those with vision impairments see the data. This is best achieved by using colors sparingly and testing their effectiveness – using color to highlight the main data is more effective than using a color for everything. Avoid relying on colors as the sole channel for conveying key information, it may help to double encode data with color and label and/or symbol (provided that this is not visually overwhelming).

- Do figures are readable in greyscale mode?
- Are colors visible to main forms of color blindness?
- Are colors used consistently across all figures?
- Avoid using colors as the sole channel for key information.
- Do not overload figures with too many colors (Magic number 7 +/-2).
- Explain all colors used.

*Background*. Tools and simulations are invaluable for verifying that your chosen colors work for all viewer; note that also visually-able audiences vary in their color perception. Most operating systems have display settings that can simulate color-blindness and greyscale, alternatively web-based tools exist. To verify that the color has a high contrast to the background consider using the WebAIM's contrast checker, aiming for a minimum contrast ratio of 4.5 to 1. More details on accessibility in visualizations for the different target groups and visual task is available (Hattab et al., 2020; Kim et al., 2021). For selecting color schemes consider tools like Paletton (https://paletton.com/) to select harmonious or contrasting color combinations that enhance the visual appeal of your work.

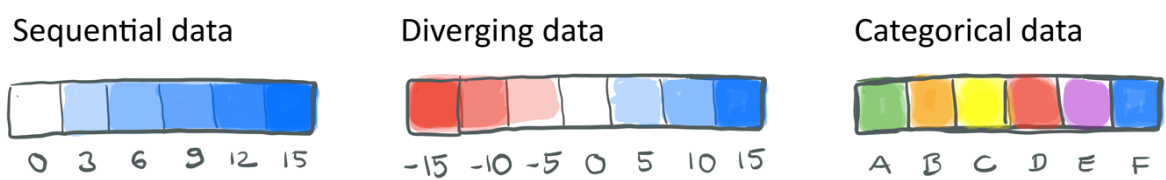

*Figure 7 Example color palettes for encoding numerical data in charts. Note, most color palettes for more than 4 categories are not color blind safe.*

## On beauty

When the charts and figures are fully functional, you may then consider making them beautiful. This can be done with an attractive color scheme (Figure 8A-B), by increasing symmetry and avoiding gaps (especially in multi-panel figures) and generally using white space to clearly separate elements (see also Focus the attention and Layout). Additionally, ensure that the charts are large enough in the final figure and in print, that the text is sized relative to the chart (examples in: Wilke, 2019), and that text size is consistent across figures and elements in multi-panel figures. A pretty way to make figures more self-explanatory is to include human recognizable icons or pictograms, that can rapidly guide the gaze in visualizations (Jambor and Bornhäuser, 2024).

- A versatile, general icon repository: https://www.svgrepo.com/
- Lab equipment and biology: https://bioicons.com/
- Plants & Animal: http://phylopic.org/
- Cells, lab stuff and animals: https://scidraw.io/

*Background*. There is neither a framework for "beauty" in data visualization, nor research whether prettier charts are more insightful or memorable. However, there are Gestalt principles that describe how our visual system makes sense from patterns (Wertheimer, 1912). We group objects when they have a similar appearance, in proximity, or are connected by lines. But we also group objects that are symmetrical to each other, which is used for instance in the organization of panel figures. Symmetry can also be applied to data by ordering bars by quantity in a bar charts, which is both attractive as well as effective to audiences. By applying symmetry to increase effectiveness of a visualization, we simultaneously make our designs attractive to audiences.

While beauty is a black box, emotion is more widely studied in visualization research. A qualitative study revealed that a positive emotion upon seeing a chart increases user engagement (Kennedy and Hill, 2018). In charts, data animations, targeted to elicit a range of emotions such as joy or amusement, increased user engagement (attention, interest, likability) to the data (Lan et al., 2022). Lastly, humans are drawn to visual items they recognize at first glance: those data visualizations that included decorative photos or pictograms, something visualization minimalists refer to as "chart junk", proved to be at least more recognizable to in an experiment with a large user group testing the recall (Borkin et al., 2013). Indeed, including or complementing chart text with suitable images does increase memorability (Haroz et al., 2015), and may even increase user engagement.

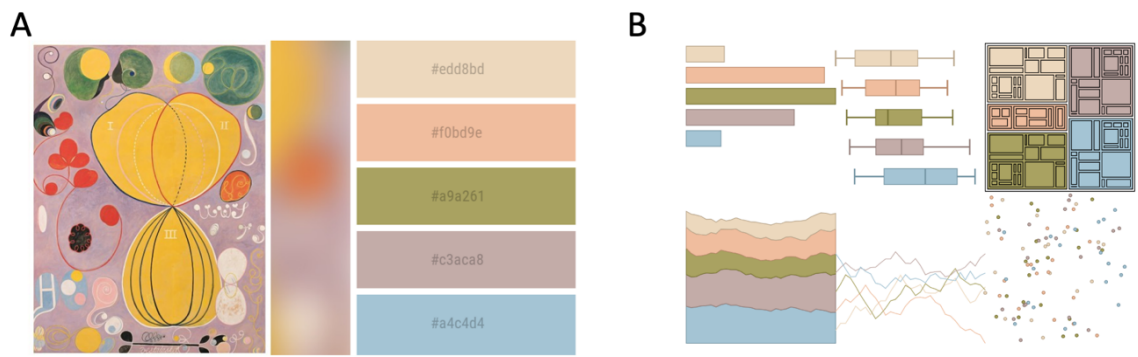

*Figure 8 A. Colors can be extracted as RGB or HEX codes from a favorite piece of art (or similar). B. The extracted colors can be tested before use with sample data and various chart types with VizPalette.*

### Bonus – the specials

**Tables**. Have a look at Table 2 and its organization. It does not only serves as a guide to further reading materials, but also itself an example of an effective table. What to consider in tables:

- Left align text (we read from left), right align numbers (we read numbers from right to left), and align each column header with its content.
- Use a font that has same width and height for all numbers (do not use proportional fonts, here 1 is smaller than 0).
- Think about the cell borders: are they needed? Also, when you remove the vertical borders, you have a little more space for text or at least to make the text more readable.

**Image data**, e.g. obtained from microscopy or photography, has a crucial role in documenting life science research. Often, this data is analyzed and summarized with charts and diagrams, however the images themselves are also published for illustration purposes in figures. The current challenges and problems of images in publications, misleading (Bik et al., 2016; Cromey, 2010) or incomprehensible figures (Jambor et al., 2021), have been described previously, as well as solutions for authors to avoid these pitfalls (North, 2006). We ourselves previously provided numerous examples of poor image figures along with re-designs (Jambor et al., 2021), a guide and cheat sheets for the design principles towards a legible image figures and for how to accomplish this with a common, open source image analysis tool (Schmied and Jambor, 2020; Senft et al., 2023), and community-guidelines informed checklists for image figures for authors to consult (Schmied et al., 2024).

**Graphical abstracts.** For advise on designing legible graphical abstracts consult the Ten Simple Rules article (Jambor and Bornhäuser, 2024) and an article about overview figures in general (Wong, 2011e).

### Conclusion

The constant development of biological research necessitates effective data visualization techniques to communicate complex and large-scale datasets clearly. In the longer run, effective data visualizations and figures aide in better data communication, but also enhance the long-term impact and memorability of scientific research. The figure checklist (Figure 1) and the accompanying explanations and resources (Table 1) provide a easy to use resource for designing legible and impactful data figures. The checklist is also a good conversation starter for those advising figure making. Complete the checklist before take-off!

### Acknowledgements

HKJ would like to acknowledge my colleagues Tracey Weissgerber, Alenka Gucek, Steve Royle and James P. Saenz for feedback on the draft version and the many students of my data-viz courses who brought their examples and questions, continuously inspiring and challenging me, and prompting me to condense my advice in this checklist.

### Financial Disclosure Statement

HKJ received a salary from an habilitation award of the Medical Faculty of the Technische Universität Dresden.

The funders had no role in study design, data collection and analysis, decision to publish, or preparation of the manuscript.

# Tables

*Table 1. Resources for making charts*

| Name | What for? | Website |
|---|---|---|
| **Chart types** | | |
| DataViz Catalogue | A collection by Severino Ribecca of chart types and their applications | http://www.datavizcatalogue.com/ |
| Python Graph Gallery | The Python Graph gallery by Yan Holz with code and application examples | https://python-graph-gallery.com/ |
| Molecular Cytology Shiny Apps | Web-applications to generate a uncommon charts, e.g. beeswarm, volcano plot, or superplots by Joachim Goedhart (Goedhart, 2021) | https://huygens.science.uva.nl/ |
| **Color** | | |
| Color Brewer by Cynthia Brewer | Color schemes to encode numerical data in charts | https://colorbrewer2.org |
| Coblis | Color blindness simulator and information about color blindness | https://www.color-blindness.com/coblis-color-blindness-simulator/ |
| Webaim | Color contrast checker | https://webaim.org/resources/contrastchecker/ |
| VizPalette | Test color schemes with example data in various chart types. By Susie Lu and Elijah Meeks | https://projects.susielu.com/viz-palette |
| Icolorpalette | Extract colors from images | https://icolorpalette.com/color-palette-from-images |
| **Icons** | | |
| BioIcons | Icons for molecular biology, from various sources, site by Simon Dürr | https://bioicons.com/ |
| Phylopic | >9000 icons of plants and animals, collected by Mike Keesey | http://phylopic.org/ |
| SciDraw | Icons of model organisms and lab equipment, collection supported by Sainsbury Wellcome Centre | https://scidraw.io/ |

*Table 2. Recommended further reading*

| Name | Author | What for? | Pages | Price |
|---|---|---|---|---|
| The digital cell | Stephen J. Royle | The book covers the entire range of data science that an experimental biologist needs for their data science – statistics, coding with R scripts, analysis, and also how to present biolgical data with plots and charts. | 137 | 67.00 |
| Fundamentals of data visualization | Claus O. Wilke | Professor for Molecular Evolution and R pacakge developer with long-standing interest in data visualization of biological data. | 387 | 49,29 |
| Truthful data, Functional data | Alberto Cairo | A data journalist revealing fundamentals about communicating data truthfully and effectively – and much of it also applies to biological data. His books are very well written. | 384 | 35.00 |
| The WSJ Guide to information graphics | Donna Wong | From the Wall Street Journal journalist Donna Wong, a very handy and concise guide. | 160 | 19.49 |
| The visual display of quantitative information | Edward Tufte | Fundamental insights in visual perception and how information design may be improved by applying these to data. | 197 | 26.82 |